\documentclass[conference]{IEEEtran}
\pdfoutput=1
\IEEEoverridecommandlockouts
\usepackage{acronym}
\usepackage{cite}
\usepackage{amsmath,amssymb,amsfonts}
\usepackage{algorithmic}
\usepackage{caption}
\usepackage{changebar}
\usepackage{changepage}
\usepackage{enumerate}
\usepackage{graphicx}
\usepackage{microtype}
\usepackage{hyperref}
\usepackage{siunitx}
\usepackage{textcomp}
\usepackage{xcolor}
\def\BibTeX{{\rm B\kern-.05em{\sc i\kern-.025em b}\kern-.08em
    T\kern-.1667em\lower.7ex\hbox{E}\kern-.125emX}}

\usepackage{amsmath, bm}
\usepackage{nicefrac}

\usepackage{tikz}
\usetikzlibrary{math}
\usetikzlibrary{shapes.geometric,calc}
\usetikzlibrary{arrows.meta, quotes, angles}
\usetikzlibrary{decorations}
\usetikzlibrary{chains,shapes.misc,positioning,scopes}
\usepackage{circuitikz}
\usetikzlibrary{circuits.ee.IEC}
\usetikzlibrary{decorations}
\usetikzlibrary{dsp}
\usetikzlibrary{shadows}

\newcommand{\dint}[1]{\,\mathrm{d}#1}
\newcommand{\Kprim}{\bm{K}}
\newcommand{\Kadj}{\tilde{\bm{K}}}

\newcommand{\tran}{^{\scriptscriptstyle\mathrm{T}}}
\newcommand{\Fbh}{\bm{F}_\mathrm{b}\tran}
\newcommand{\As}{\bm{\mathcal{A}}}
\newcommand{\Cs}{\bm{\mathcal{C}}}
\newcommand{\Csa}{\tilde{\bm{\mathcal{C}}}}
\newcommand{\Ks}{\tilde{\bm{\mathcal{K}}}}

\acrodef{DMC}{Diffusive Molecular Communication}
\acrodef{ISI}{inter-symbol interference}
\acrodef{MC}{Molecular Communication}
\acrodef{NT}{neurotransmitter}
\acrodef{wrt}[w.r.t.]{with respect to}
\acrodef{SSD}{state-space description}
\acrodef{PBS}{particle-based simulations}

\newcommand{\dtau}{\ensuremath{\mathrm{d}\tau}}
\newcommand{\ka}{\ensuremath{\kappa_a}}
\newcommand{\kd}{\ensuremath{\kappa_d}}
\newcommand{\kad}{\ensuremath{\kappa_{a_0}}}
\newcommand{\ihat}{\ensuremath{\hat{i}(t)}}
\newcommand{\xmin}{\ensuremath{x_{\mathrm{min}}}}
\newcommand{\xmax}{\ensuremath{x_{\mathrm{max}}}}
\newcommand{\Omegax}{\ensuremath{\Omega_{\textrm{x}}}}
\newcommand{\kece}{\ensuremath{\kappa_e C_E}}

\IEEEaftertitletext{\vspace{-1.2\baselineskip}}

\renewcommand{\baselinestretch}{0.97}

\makeatletter
\long\def\@makecaption#1#2{\ifx\@captype\@IEEEtablestring%
    \footnotesize\begin{center}{\normalfont\footnotesize #1}\\
        {\normalfont\footnotesize\scshape #2}\end{center}%
    \@IEEEtablecaptionsepspace
    \else
    \@IEEEfigurecaptionsepspace
    \setbox\@tempboxa\hbox{\normalfont\footnotesize {#1.}~~ #2}%
    \ifdim \wd\@tempboxa >\hsize%
    \setbox\@tempboxa\hbox{\normalfont\footnotesize {#1.}~~ }%
    \parbox[t]{\hsize}{\normalfont\footnotesize \noindent\unhbox\@tempboxa#2}%
    \else
    \hbox to\hsize{\normalfont\footnotesize\hfil\box\@tempboxa\hfil}\fi\fi}
\makeatother

\makeatletter
\renewcommand\@afterheading{%
    \@nobreaktrue
    \everypar{%
        \if@nobreak
        \@nobreakfalse
        \clubpenalty 1
        \if@afterindent \else
        {\setbox\z@\lastbox}%
        \fi
        \else
        \clubpenalty 1
        \everypar{}%
        \fi}}
\makeatother

\begin{document}

\title{Receptor Saturation Modeling for Synaptic DMC
\thanks{This work was supported in part by the German Research Foundation (DFG) under grants SCHO 831/9-1 and SCHO 831/14-1.}
\vspace{-2mm}
}

\author{\IEEEauthorblockN{Sebastian Lotter, Maximilian Sch\"afer, Johannes Zeitler, and Robert Schober} 
    \IEEEauthorblockA{Friedrich-Alexander University Erlangen-Nuremberg, Germany}
}
\maketitle
\begin{abstract}
Synaptic communication is a natural \ac{MC} system which may serve as a blueprint for the design of synthetic \ac{MC} systems.
In particular, it features highly specialized mechanisms to enable \ac{ISI}-free and energy efficient communication.
The understanding of synaptic \ac{MC} is furthermore critical for disruptive innovations in the context of brain-machine interfaces.
However, the physical modeling of synaptic \ac{MC} is complicated by the possible saturation of the molecular receiver arising from the competition of postsynaptic receptors for neurotransmitters.
Saturation renders the system behavior nonlinear and is commonly neglected in existing analytical models.
In this work, we propose a novel model for receptor saturation in terms of a nonlinear, state-dependent boundary condition for Fick's diffusion equation.
We solve the resulting boundary-value problem using an eigenfunction expansion of the Laplace operator and the incorporation of the receiver memory as feedback system into the corresponding state-space description.
The presented solution is numerically stable and computationally efficient.
Furthermore, the proposed model is validated with particle-based stochastic computer simulations.
\end{abstract}

\acresetall
\section{Introduction}
\ac{MC} is a bio-inspired communication paradigm in which information is transmitted by molecules.
It has gained significant attention as potential enabler of novel applications in the context of the internet of Bio-nano things \cite{akyildiz15}.
In particular, \ac{MC} is considered as promising candidate for novel intra-body applications due to its inherent bio-compatibility and the fact that traditional electromagnetic wave-based wireless communication is not feasible at the nano-scale \cite{nakano13}.
Natural \ac{MC} systems evolved over millions of years to cope with the challenges faced in intra-body nano-scale communication and might hence serve as blueprints for synthetic \ac{MC} systems.
Among the different natural types of \ac{MC}, \ac{DMC}, i.e., communication via diffusing molecules, is a promising candidate for synthetic \ac{MC} as it requires neither dedicated communication infrastructure nor external energy sources for molecule propagation \cite{nakano13}.

\begin{figure}[!t]
    \centering
    \includegraphics[width=.45\textwidth]{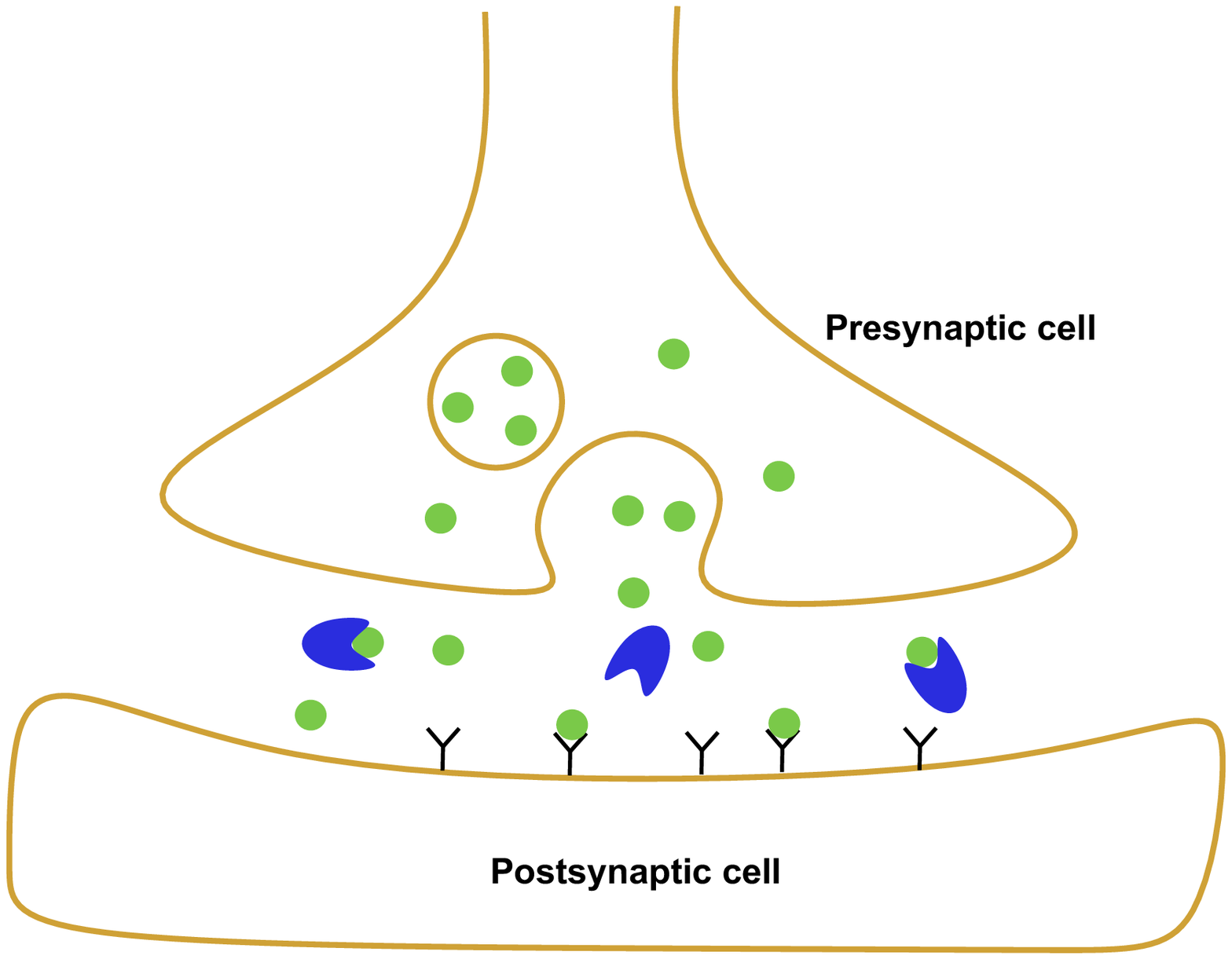}\vspace*{-3 mm}
    \caption{Model synapse. Neurotransmitters (green) enclosed in vesicles are released at the presynaptic cell, propagate by Brownian motion, and activate receptors at the postsynaptic cell. Binding to postsynaptic receptors is reversible. Solute neurotransmitters are degraded by enzymes (blue) \cite{zucker14}.}\vspace*{-6 mm}
    \label{fig:channel}
\end{figure}

\acp{DMC} can be found in the human body for example in chemical synapses formed between adjacent nerve cells or between neurons and muscle fibers.
The synaptic communication system comprises in its simplest form the presynaptic cell (transmitter), the postsynaptic cell (receiver), and the synaptic cleft (channel) \cite{hof14}, cf.~Fig.~\ref{fig:channel}.
The message carrying molecules in this highly specialized \ac{DMC} system are termed \acp{NT}.
To convey information, \acp{NT} are released from vesicular containers at the presynaptic cell, diffuse across the synaptic cleft and bind to postsynaptic receptors \cite{nakano13}.
The signal is terminated by molecule uptake or enzymatic degradation \cite{nakano13}.

One of the challenges for modeling the synaptic \ac{DMC} system is that a finite number of postsynaptic receptors compete for \acp{NT} rendering the system under consideration nonlinear in the number of released \acp{NT}.
This effect is called {\em receptor saturation} and significantly impacts synaptic transmission at some types of synapses \cite{foster05}.
Hence, in particular for the design of synthetic \ac{DMC} systems, it is desirable to understand how and under which conditions receptor saturation impacts signal transmission in synaptic \ac{DMC}.

Synaptic communication has been studied in the \ac{MC} literature before and we refer the reader to \cite{veletic2019,lotter20} for recent literature overviews.
In most models, e.g.~in \cite{balevi13,liu14,veletic16b,lotter20a}, postsynaptic receptor saturation is neglected.
In \cite{khan2017}, a finite number of postsynaptic receptors is assumed, but the impact of \ac{NT} buffering at postsynaptic receptors on the concentration of solute \acp{NT} is not taken into account. 
This approach leads potentially to an underestimation of the synaptic \ac{ISI} caused by residual \acp{NT} in the synaptic cleft.
Receptor saturation and its impact on the concentration of solute \acp{NT} in the presence of presynaptic \ac{NT} transporters is modeled in \cite{bilgin17}.
The resulting nonlinear model is solved by dicretizing the diffusion equation in space and time and employing an iterative numerical algorithm.
In \cite{veletic19a}, a non-linear model for ligand-receptor binding based on the chemical master equation is analyzed using Volterra series.
The spatial distribution of molecules is, however, not considered in \cite{veletic19a}.

Saturation at the molecular receiver has been considered in the \ac{MC} literature also in the context of targeted drug delivery \cite{femminella15,felicetti16,salehi19} and experimental studies \cite{farsad14,kim15,kim19}.
In none of these works, however, the impact of receptor saturation on the spatial distribution of solute molecules is modeled explicitly.
Finally, in \cite{ahmadzadeh16}, receptor saturation is studied in an unbounded environment using \ac{PBS}.

In this work, we present a novel model for synaptic \ac{DMC} incorporating receptor saturation and enzymatic molecule degradation.
The proposed model is based on the diffusion equation and incorporates an analytical model of the reversible binding of \acp{NT} to a finite number of postsynaptic receptors in terms of a saturation boundary condition.
In contrast to previous works, our model encompasses a spatial model of the synaptic cleft and a finite number of postsynaptic receptors without decoupling the concentrations of solute and bound molecules (as e.g.~in \cite{khan2017}) or the need for spatial discretization (as e.g.~in \cite{bilgin17}).
Our approach exploits the modeling of the diffusion equation in terms of a \ac{SSD} \cite{schaefer:ecc:2019}.
It is based on a functional transformation of the diffusion equation adapted to the synaptic geometry and allows the modular incorporation of the nonlinear receptor saturation effect by a feedback structure \cite{schaefer:laminar:2020}.
Compared to particle-based Monte Carlo methods, the approach presented in this paper is computationally extremely efficient as the computational cost scales neither with the number of released particles nor with the number of receptors.
Furthermore, it yields the expected received signal without diffusion noise.
The proposed model is to the best of the authors' knowledge the first spatial analytical model to simultaneously consider enzymatic degradation and receptor saturation.
Results from \ac{PBS} validate the accuracy of our model.

The remainder of this paper is organized as follows:
In Section \ref{sec:system_model}, the system model is introduced.
In Sections \ref{sec:transfer_function_model} and \ref{sec:state-space_description}, we develop the proposed transfer function model and the resulting \ac{SSD}, respectively.
The results are presented and compared with \ac{PBS} in Section \ref{sec:results}, and the main findings are summarized in Section \ref{sec:conclusion}.

\section{System Model}\label{sec:system_model}
\subsection{Assumptions}\label{sec:system_model:geometry_and_assumptions}
The shapes of natural synapses are highly variable \cite{hof14}. 
In this work, we adopt the cuboid model for the synaptic cleft proposed previously in \cite{lotter20a}.
Formally, it is defined in Cartesian coordinates as follows \cite{lotter20a}
\begin{multline}
\Omega = \{(x,y,z) \vert x_{\mathrm{min}} \leq x \leq x_{\mathrm{max}},\\ y_{\mathrm{min}} \leq y \leq y_{\mathrm{max}}, z_{\mathrm{min}} \leq z \leq z_{\mathrm{max}} \}.\label{eq:domain}
\end{multline}
In this model, the faces $x=\xmin$ and $x=\xmax$ represent the membranes of the presynaptic and postsynaptic neuron, respectively, and the faces in $y$ and $z$ are reflective and constrain the synaptic cleft.
After \acp{NT} are released from presynaptic vesicles \cite{zucker14}, they are in nature either uptaken by the presynaptic neuron or surrounding glial cells or degraded by enzymes to terminate synaptic signaling \cite{nakano13}.
While presynaptic and glial cell uptake has been considered previously by the authors (without receptor saturation) \cite{lotter20a,lotter20}, in this work, we focus on enzymatic degradation as clearance mechanism.

We define the postsynaptic signal as the {\em number of bound (postsynaptic) receptors at time $t$}, $\ihat$.
In the following, we are interested in the impact of receptor saturation on $\ihat$ in response to a single release or multiple releases of \acp{NT}.

Before we state the system model, we introduce the following assumptions:
\begin{enumerate}[{A}1)]
    \item The diffusion coefficient and the total number of postsynaptic receptors are time-invariant over the time frame under consideration.\label{ass:const_rates}
    \item The diffusive propagation of molecules is fast relative to the binding to postsynaptic receptors.\label{ass:slow_reactions}
    \item Enzymatic degradation can be modeled as first-order reaction and the enzymes are uniformly distributed in space.\label{ass:enz_deg}
    \item Reversible adsorption to individual, uniformly distributed receptors with intrinsic association coefficient $\kad$ in $\si{\micro\meter\per\micro\second}$ and intrinsic dissociation rate $\kd$ in $\si{\per\micro\second}$ can be treated equivalently as reversible adsorption to a homogeneous surface with effective association coefficient $\ka$ in $\si{\micro\meter\per\micro\second}$ and dissociation rate $\kd$.\label{ass:bdr_hgz}
\end{enumerate}

A\ref{ass:const_rates} is justified if the time frame under consideration is sufficiently small as the insertion and removal of postsynaptic receptors constitutes a long-term adaptation process \cite{gallimore16}.

A\ref{ass:slow_reactions} is plausible given experimentally observed values for the diffusion coefficient of the common \ac{NT} glutamate \cite{nielsen04} and postsynaptic receptor binding rates \cite{jonas93}.
A\ref{ass:slow_reactions} guarantees that the \acp{NT} are approximately uniformly distributed in $y$ and $z$ and, hence, the \ac{NT} concentration in $\Omega$ can be equivalently characterized by the one-dimensional \ac{NT} concentration in $\Omegax = [\xmin;\xmax]$.
We note that, because of the reflective boundaries in $y$ and $z$, $\Omega$ would in fact be equivalent to $\Omegax$ in terms of $\ihat$, if receptors did not saturate \cite{lotter20a}.
The validity of this assumption is further investigated in Section~\ref{sec:results} by comparing the results of three-dimensional \ac{PBS} with the proposed one-dimensional model.

The first part of A\ref{ass:enz_deg} is justified in \cite{noel14} under the assumption that the degradation of molecules bound to enzymes is sufficiently fast.
The latter part is reasonable given that the enzymes exist long enough such that their concentration reaches equilibrium.
As a consequence of A\ref{ass:enz_deg}, the time constant of enzymatic degradation is given as $\kappa_e C_E$, where $C_E$ denotes the constant concentration of degrading enzymes in \si{\per\micro\meter} and $\kappa_e$ denotes the degradation rate in \si{\micro\meter\per\micro\second} \cite{noel14}.

A\ref{ass:bdr_hgz} is referred to as {\em boundary homogenization} and has been investigated in \cite{lotter20a}.
The validity of A\ref{ass:bdr_hgz} is further confirmed by the results presented in Section~\ref{sec:results}.

With these assumptions, we now formulate the analytical system model.

\subsection{Synaptic Neurotransmitter Concentration}\label{sec:system_model:nt_concentration}
Using assumption A\ref{ass:slow_reactions} from Section~\ref{sec:system_model:geometry_and_assumptions} and setting $\xmin = 0$ and $\xmax = a$, the concentration of \acp{NT} in the synaptic cleft is governed by the inhomogeneous reaction-diffusion equation
\begin{equation}
\partial_t c(x,t) = D\partial_{xx} c(x,t) - \kece c(x,t) + q(x,t), \quad 0 < x < a,\label{eq:diff_oned}
\end{equation}
where $c(x,t)$ denotes the \ac{NT} concentration in \si{\per\micro\meter}, $D$ is the diffusion coefficient in \si{\micro\meter\squared\per\micro\second}, and $\partial_t$ and $\partial_{xx}$ denote the first derivative \ac{wrt} time and the second derivative \ac{wrt} space, respectively.
The term $q(x,t)$ in \eqref{eq:diff_oned} is independent of $c(x,t)$ and models the release of \acp{NT} into the synaptic cleft (in \si{\per\micro\meter\per\micro\second}).

The boundary condition at the presynaptic membrane, $x=0$, is given by the no-flux boundary condition
\begin{equation}
    - D\,\partial_x c(x,t)\big\vert_{x=0} = i(0,t) = 0, \label{eq:pab_lb}
\end{equation}
where $\partial_x$ denote the first derivative \ac{wrt} space and $i(x,t)$ denotes the particle flux in \si{\per\micro\second}.

If the receptors did not saturate, the binding of \acp{NT} to postsynaptic receptors could be modeled as reversible adsorption to a homogeneous, partially absorbing boundary \cite{lotter20a}.
The corresponding boundary condition would then be \cite{lotter20a}
\begin{equation}
i(a,t) = \ka c(a,t) - \kd \ihat,\label{eq:pab_rb_no_sat}
\end{equation}
where $\ka$ denotes the effective adsorption coefficient in \si{\micro\meter\per\micro\second} resulting from homogenizing the postsynaptic boundary, $\kd$ denotes the dissociation rate in \si{\per\micro\second},
\begin{equation}
\hat{i}(t) = \int_{0}^{t} i(a,\tau) \dtau,\label{eq:def:ihat}
\end{equation}
and we have assumed that $\hat{i}(0) = 0$.

Now, the saturation of postsynaptic receptors introduces memory to the adsorption process in the sense that the rate of adsorption at time $t=t_1$ depends on how many receptors are occupied at time $t_1$, i.e., $\hat{i}(t_1)$, which in turn depends on the entire history of binding and unbinding of \acp{NT} to receptors.

Using \eqref{eq:def:ihat}, we propose to incorporate saturation into \eqref{eq:pab_rb_no_sat} as follows
\begin{equation}
i(a,t) = \ka \left( 1 - \nicefrac{\hat{i}(t)}{C^*} \right) c(a,t) - \kd \hat{i}(t),\label{eq:pab_rb}
\end{equation}
where $C^*$ denotes the total number of postsynaptic receptors.
Considering the term $\left( 1 - \nicefrac{\hat{i}(t)}{C^*} \right)$ in \eqref{eq:pab_rb}, molecules bind with the full rate $\ka$ if no receptors are occupied, i.e., $\hat{i}(t)=0$, and binding drops to zero, if all receptors are occupied, i.e., $\hat{i}(t)=C^*$.
As $\hat{i}(t)$ depends on the current and all past values of $c(a,t)$, \eqref{eq:pab_rb} is a nonlinear, state-dependent boundary condition which we term {\em saturation boundary condition}.

To complete the formulation of the model, we require that the initial concentration of \acp{NT} in the synaptic cleft is zero at $t=0$, i.e., $c(x,0)=0$.

\section{Transfer Function Model}\label{sec:transfer_function_model}
In this section, we formulate the boundary-value problem from Section~\ref{sec:system_model:nt_concentration} in terms of transfer functions.
We first consider the special case $\ka=\kd=\kece=0$.
This model is then extended in Section~\ref{sec:state-space_description} to the general case $\ka,\kd,\kece \geq 0$.

\subsection{Vector Representation}
Assuming $\kece=0$, \eqref{eq:diff_oned} is decomposed into two equations and arranged into vector form as follows \cite[Eq.~(12)]{schaefer:laminar:2020}
\begin{align}
	&\left[\partial_t\bm{D} - \bm{\mathrm{L}}\right]\bm{y}(x,t) = \bm{f}(x,t), \label{eq:tf:1}
\end{align}
where
\vspace*{-0.5ex}
\begin{align}
	\bm{D} &= \begin{bmatrix}
		0 & 0\\
		1 & 0
	\end{bmatrix}, 
	&\bm{\mathrm{L}} = \begin{bmatrix}
		- \partial_x & -\nicefrac{1}{D} \\
		0 & -\partial_x
	\end{bmatrix},
	\label{eq:tf:2}\\
	\bm{y}(x,t) &= 
	\begin{bmatrix}
		c(x,t)&
		i(x,t)
	\end{bmatrix}\tran\!\!, 
	&\bm{f}(x,t) = 
	\begin{bmatrix}
		0&
		q(x,t)
	\end{bmatrix}\tran,
	\label{eq:tf:3}
\end{align}
and $(\cdot)\tran$ denotes transposition.
Eqs.~\eqref{eq:pab_lb}, \eqref{eq:pab_rb} are represented with the boundary operator $\Fbh\in\mathbb{R}^{2\times2}$ acting on $\bm{y}(x,t)$ as
\begin{align}
	&\Fbh\bm{y}(x,t) = \bm{\phi}(x,t), &x = 0,a,
	\label{eq:tf:4}
\end{align}
where $\mathbb{R}$ denotes the set of real numbers, and $\Fbh$ and the vectorized boundary values $\bm{\phi}(0,t)$ and $\bm{\phi}(a,t)$ are defined as follows
\begin{align}
	&\Fbh = \begin{bmatrix}
		0 & 0\\
		0 & 1
	\end{bmatrix},
	& \bm{\phi}(0,t) = \bm{0},
	&&\bm{\phi}(a,t) = \begin{bmatrix}
		0\\
		p(t)
	\end{bmatrix}.
	\label{eq:tf:5}
\end{align}
The time-variant boundary value $p(t)$ in \eqref{eq:tf:5} is used as placeholder for the right-hand side of \eqref{eq:pab_rb} and we assume for the moment that it is independent of $\bm{y}(x,t)$.
Eqs.~\eqref{eq:tf:1}, \eqref{eq:tf:4} describe a one-dimensional diffusion process with Neumann boundary conditions.

\subsection{Functional Transformations}

The solution of \eqref{eq:tf:1}, \eqref{eq:tf:4}, $\bm{y}(x,t)$, is expanded into the infinite set of bi-orthogonal eigenfunctions $\Kprim_\mu(x) \in\mathbb{R}^{2\times 1}$ and $\Kadj_\mu(x) \in\mathbb{R}^{2\times 1}$ of the spatial differentiation operator $\bm{\mathrm{L}}$.
The corresponding eigenvalues $s_\mu$ define the discrete spectrum of $\bm{\mathrm{L}}$ \cite{churchill:1972}.
Both, eigenvalues and eigenfunctions are indexed with $\mu\in\mathbb{N}_0$, where $\mathbb{N}_0$ denotes the set of non-negative integers.
Due to space constraints, the spatial transformations required for the expansion of \eqref{eq:tf:1} are not presented here and we refer the reader to \cite[Secs.~III, IV]{schaefer:ecc:2019} for a detailed exposition.
Instead, only the parts necessary for the construction of the solution of \eqref{eq:tf:1} are provided.

The eigenfunctions $\Kprim_\mu(x)$ and $\Kadj_\mu(x)$ are derived using the procedure in \cite[Sec.~III]{schaefer:nds:2017a} as
\begin{align}
	&\Kprim_\mu(x) \!=\! \begin{bmatrix}
		\cos(\gamma_\mu x)\\
		D\gamma_\mu\sin(\gamma_\mu x)
	\end{bmatrix}\!\!,\!\!\!\!
	&\Kadj_\mu(x) \!=\! \begin{bmatrix}
		-D \gamma_\mu \sin(\gamma_\mu x)\\
		\cos(\gamma_\mu x) 
	\end{bmatrix}\!\!.
	\label{eq:tf:6} 
\end{align}
The eigenvalues $s_\mu$ and wavenumbers $\gamma_\mu$ can be derived from \eqref{eq:tf:4} as $s_\mu = -D\,\gamma_\mu^2$, $\gamma_\mu = \mu \frac{\pi}{a}$ \cite[Sec.~IV]{schaefer:nds:2017a}.
The eigenfunctions in \eqref{eq:tf:6} have to be bi-orthogonal to ensure the existence of an inverse transformation \cite{churchill:1972}, yielding the factor 
\begin{align}
	N_\mu = \int_0^a\Kadj_\mu\tran(x)\bm{D}\Kprim_\mu(x)\dint{x} = \begin{cases}
		a & \mu = 0\\
		\nicefrac{a}{2} & \mu \neq 0
	\end{cases}.
	\label{eq:tf:8}
\end{align}
Finally, $\bm{y}(x,t)$ is given by the following series expansion
\begin{align}
	\bm{y}(x,t) &= \sum_{\mu = 0}^{\infty} \frac{1}{N_\mu}\bar{y}_\mu(t) \Kprim_\mu(x), \label{eq:tf:9}
\end{align}
with the expansion coefficients
\begin{align}
	\bar{y}_\mu(t) &= \mathrm{e}^{s_\mu t} \overset{t}{*}\left(\bar{f}_\mu(t) - \bar{\phi}_\mu(t)\right).\label{eq:tf:10}
\end{align}
Here, $\overset{t}{*}$ denotes the convolution \ac{wrt} time, and $\bar{f}_\mu(t)$ and $\bar{\phi}_\mu(t)$ follow from $\bm{f}(x,t)$ in \eqref{eq:tf:3} and $\phi(x,t)$ in \eqref{eq:tf:5} as
\begin{align}
	&\bar{f}_\mu(t) \!=\!\!\int_0^a\!\!\!\Kadj_\mu\tran(x)\bm{f}(x,t)\dint{x},\!\!\!
	&\bar{\phi}_\mu(t) \!=\! \left[\Kadj_\mu\tran(x) \bm{\phi}(x,t)\right]_0^a\!\!\!\quad, \label{eq:tf:11}
\end{align}
where $[f(x)]_a^b = f(b) - f(a)$.
For the numerical evaluation, the infinite sum in \eqref{eq:tf:9} is truncated to $\mu = 0, \dots, Q-1$.
The accuracy of the computed solution hence depends on $Q$.

\newpage
\changepage{}{}{}{}{}{0.02in}{}{}{}
\section{State-space description}
\label{sec:state-space_description}
In this section, we consider the general case $\ka,\kd,\kece \geq 0$.
To this end, the model in \eqref{eq:tf:9} is first transformed into the discrete-time using an impulse invariant transform \cite{schaefer:laminar:2020}.
This yields the following discrete-time \ac{SSD} \cite{schaefer:ecc:2019}
\begin{align}
	\bar{\bm{y}}[k+1] &= \mathrm{e}^{\As T}\bar{\bm{y}}[k] +T\bar{\bm{f}}[k+1] - T\bar{\bm{\phi}}[k+1], \label{eq:tf:12}\\
	\bm{y}[x,k] &= \Cs(x)\bar{\bm{y}}[k], \label{eq:tf:13}
\end{align}
with discrete-time index $k$ and sampling interval $T$, i.e., $t = kT$.
Consequently, $\bm{y}[x,k]$, $c[x,k]$, and $i[x,k]$ are given by $\bm{y}(x,kT)$, $c(x,kT)$, and $i(x,kT)$, respectively.
$T$ should be adapted to the smoothness of $\bm{y}(x,t)$ to ensure that $\bm{y}(x,t)$ is accurately reproduced by $\bm{y}[x,k]$; i.e., the smaller $D$, $\kad$, $\kd$, and $\kece$ are, the smoother signal $\bm{y}(x,t)$ is and the larger may $T$ be chosen.
For a numerical example, please see Table~\ref{tab:sim_params}.
Here, \textit{state equation} \eqref{eq:tf:12} is the vector-valued discrete-time equivalent of \eqref{eq:tf:10}, where vector $\bar{\bm{y}}\in\mathbb{R}^{Q\times 1}$ contains $Q$ coefficients $\bar{y}_\mu$ and diagonal matrix $\As\in\mathbb{R}^{Q\times Q}$ contains $Q$ eigenvalues $s_\mu$ on its main diagonal, i.e., $\bar{\bm{y}}[k]=[\bar{y}_{0}(kT),\ldots,\bar{y}_{Q-1}(kT)]\tran=\left(\bar{y}_{\mu}(kT)\right)_{\mu=0}^{Q-1}$ and $\As \!=\! \mathrm{diag}\left\lbrace s_0, \dots, s_{Q-1}\right\rbrace$.
Vectors $\bar{\bm{f}}[k]\in\mathbb{R}^{Q\times 1}$ and $\bar{\bm{\phi}}[k]\in\mathbb{R}^{Q\times 1}$ are defined as $\bar{\bm{f}}[k] = \left(\bar{f}_{\mu}(kT)\right)_{\mu=0}^{Q-1}$ and $\bar{\bm{\phi}}[k] = \left(\bar{\phi}_{\mu}(kT)\right)_{\mu=0}^{Q-1}$, respectively.
\textit{Output equation} \eqref{eq:tf:13} is the discrete-time equivalent of \eqref{eq:tf:9}, where the summation is replaced by a multiplication with matrix $\Cs(x)\in\mathbb{R}^{2\times Q}$,
\begin{align}
\Cs(x) &= \left[\nicefrac{1}{N_0}\Kprim_0(x), \dots, \nicefrac{1}{N_{Q-1}}\Kprim_{Q-1}(x) \right].\label{eq:tf:15}
\end{align}
For the following steps, we further define matrix $\tilde{\Cs}(x)\in\mathbb{R}^{2\times Q}$,
\begin{align}
	\Csa(x) &= \left[\Kadj_0(x), \dots, \Kadj_{Q-1}(x) \right].	\label{eq:tf:16}
\end{align}

\subsection{Incorporation of Receptor Saturation}
\label{subsec:sat}

Now, we expand the placeholder boundary value $p(t)$ introduced in \eqref{eq:tf:5} in the discrete-time domain as
\begin{align}
	p[k+1] = \hat{\kappa}_a[k]\, c[a,k] - \hat{\kappa}_d[k], \label{eq:tf:17}
\end{align}
where
\begin{align}
	&\hat{\kappa}_a[k] = \kappa_a\left(1 - \nicefrac{\hat{i}[k]}{C^*}\right), 
	&\hat{\kappa}_d[k] = \kappa_d\,\hat{i}[k],  \label{eq:tf:18}
\end{align}
and the discrete-time accumulated net flux, $\hat{i}[k]$, is defined as
\begin{align}
	\hat{i}[k] = T\sum_{n = 0}^{k}i[a,n]. \label{eq:tf:19}
\end{align}
In \eqref{eq:tf:17}, the value of $p$ in time slot $k+1$ depends on the values of the concentration $c$ and flux $\hat{i}$ in time slot $k$. 
Hence, \eqref{eq:tf:17} introduces a delay of $T$ in the computation of the flux compared to the right-hand side of \eqref{eq:pab_rb}.
However, if $T$ is chosen small enough relative to the velocity of the binding kinetics defined by $\kappa_a$ and $\kappa_d$, this simplification is justified because $\hat{i}$ and $c$ will be approximately constant in two subsequent sampling intervals.
We verify the accuracy of this assumption in Section~\ref{sec:results} with \ac{PBS}.

Now, vector $\bar{\bm{\phi}}$ in \eqref{eq:tf:12} can be evaluated using the vector-valued discrete-time version of \eqref{eq:tf:11}
\begin{align}
	\bar{\bm{\phi}}[k+1] &= \left[\Csa\tran(x) \bm{\phi}(x,(k+1)T)\right]_0^a. \label{eq:tf:20}
\end{align}
Exploiting the structure of $\bm{\phi}(x,t)$ in \eqref{eq:tf:5}, $\tilde{\Cs}(x)$ in \eqref{eq:tf:16}, and the definition of $p$ in \eqref{eq:tf:17}, we can write \eqref{eq:tf:20} as
\begin{align}	
	\bar{\bm{\phi}}[k+1] &=\tilde{\bm{c}}_2(a)p[k+1]\nonumber\\
	&= \tilde{\bm{c}}_2(a)\hat{\kappa}_a[k]\, c[a,k] - \tilde{\bm{c}}_2(a)\hat{\kappa}_d[k], \label{eq:tf:21}
\end{align}
where $\tilde{\bm{c}}_2(x)\in\mathbb{R}^{Q\times 1}$ is the second column of $\Csa\tran(x)$ containing the second entries of $\Kadj$ in \eqref{eq:tf:6}.
Furthermore, concentration $c$ can be expressed as follows
\begin{align}
	c[a,k] = \bm{c}_1\tran(a)\bar{\bm{y}}[k], \label{eq:tf:22}
\end{align}
where $\bm{c}_1\tran(x)\in\mathbb{R}^{1\times Q}$ is the first row of matrix $\Cs(x)$ in \eqref{eq:tf:15}.
Inserting \eqref{eq:tf:22} into \eqref{eq:tf:21} leads to
\begin{align}
	\bar{\bm{\phi}}[k+1] &= \hat{\kappa}_a[k]\Ks_a\bar{\bm{y}}[k] -\hat{\kappa}_d[k] \Ks_d, \label{eq:tf:23}
\end{align}
where $\Ks_a = \tilde{\bm{c}}_2(a)\bm{c}_1(a)$, $\Ks_d = \tilde{\bm{c}}_2(a)$.
Inserting \eqref{eq:tf:23} into \eqref{eq:tf:12}, we obtain the state equation for $\ka,\kd \geq 0$ as
\begin{align}
	\bar{\bm{y}}[k+1] &= \left(\mathrm{e}^{\As T} - T\hat{\kappa}_a[k]\Ks_a\right)\bar{\bm{y}}[k] \nonumber  \\  &\quad\quad\quad + T\hat{\kappa}_d[k]\Ks_d+ T\bar{\bm{f}}[k+1]. \label{eq:tf:25}
\end{align}

\subsection{Incorporation of Degradation}

To complete the model, the enzymatic degradation in \eqref{eq:diff_oned} has to be reincorporated into \eqref{eq:tf:25}.
As the degradation reaction is modeled as first-order reaction, it can be incorporated into \eqref{eq:tf:9} by a decaying exponential function $\mathrm{e}^{-\kece t}$ \cite{noel14}, yielding the following discrete-time model
\vspace*{-1mm}
\begin{align}
	\bar{\bm{y}}[k+1] &= \left(\mathrm{e}^{-\kece T}\mathrm{e}^{\As T} - T\hat{\kappa}_a[k]\Ks_a\right)\bar{\bm{y}}[k] \nonumber \\  &\quad\quad\quad+ T\hat{\kappa}_d[k] \Ks_d + T\bar{\bm{f}}[k+1]. \label{eq:tf:26} 
\end{align}
This modified state equation accounts for saturation and desorption at $x = a$ according to \eqref{eq:pab_rb} and enzymatic degradation, while the output equation \eqref{eq:tf:13} to calculate the NT concentration and flux remains unchanged.
We note that \eqref{eq:tf:26} collapses to \eqref{eq:tf:12} if $\ka=\kd=\kece=0$.

\section{Results}\label{sec:results}
\subsection{Particle-based Simulation}

To verify the accuracy of the state-space model derived in Section~\ref{sec:state-space_description}, three-dimensional \ac{PBS} were conducted.
To this end, the simulator design previously presented in \cite{lotter20a} was extended to account for receptor saturation and enzymatic degradation.
Receptor saturation was incorporated into the simulator by setting the binding probability for a receptor to zero when a molecule was bound to this receptor, and back to its original value when the molecule unbound.
Enzymatic degradation was incorporated by introducing a first-order degradation step for all solute molecules with probability \cite{andrews09} $1 - \exp(-\kappa_e C_E \Delta t)$, where $\Delta t$ denotes the simulation time step in \si{\micro\second}.
The results presented in the following subsections were obtained for $\Delta t = 10^{-2}$~$\si{\micro\second}$ and averaged over $50$ simulation runs.

The computational cost of the \ac{PBS} scales with the simulation time step as well as with the number of released particles, the number of receptors, and the number of simulation runs.
The runtime of the proposed \ac{SSD} model, in contrast, scales only with the sampling interval and the number of eigenfunctions.
Consequently, for the parameter values considered in this paper, the computation of the \ac{SSD} model required far less (by more than a factor of $100$) CPU time than the \ac{PBS}.

In Sections~\ref{sec:results:steady_state} and \ref{sec:results:single_release}, we set $q(x,t)=N\delta(x)\delta(t)$, where $N$ denotes the number of released molecules and $\delta(\cdot)$ denotes the Dirac delta function.
In Section~\ref{sec:results:multiple_releases}, we set $q(x,t)=N\delta(x)(\delta(t) + \delta(t - \SI{1}{\milli\second}) + \delta(t - \SI{2}{\milli\second}))$.
If not indicated otherwise, the parameter values listed in Table~\ref{tab:sim_params} were used.

\begin{table}
    \vspace*{0.07in}
    \centering
    \caption{Simulation parameters for particle-based simulation \cite{lotter20}.}
    \vspace*{-0.02in}
    \footnotesize
    \begin{tabular}{| p{.16\linewidth} | r | p{.38\linewidth} |}
        \hline Parameter & Default Value & Description\\ \hline
        $D$ & $\SI{3.3e-4}{\micro\meter\squared\per\micro\second}$& Diffusion coefficient\\ \hline
        $N$ & $\SI{1000}{}$ & Number of released particles\\ \hline
        $a$ & $2 \times 10^{-2}$~$\si{\micro\meter}$& Channel width in $x$\\ \hline
        $\{y,z\}_{\textrm{max}}-\{y,z\}_{\textrm{min}}$ & $\SI{0.15}{\micro\meter}$ & Channel widths in $y$ and $z$\\ \hline
        $\kad$ &$1.02 \times 10^{-4}$~$\si{\micro\meter\per\micro\second}$ & Intrinsic binding rate\\ \hline
        $\kd$ &$8.5 \times 10^{-3}$~$\si{\per\micro\second}$ & Intrinsic unbinding rate\\ \hline
        $\kappa_e C_{E}$ &$10^{-3}$~$\si{\per\micro\second}$ & Degradation rate\\ \hline
        $r$ & $2.3 \times 10^{-3}$~$\si{\micro\meter}$ & Receptor radius\\ \hline
        $C^*$ & $203$ & Number of uniformly distribu\-ted receptors (15\% coverage)\\ \hline
        $\Delta t$ & $10^{-2}$~$\si{\micro\second}$ & Simulation time step\\ \hline
        $Q$ & $100$ & Number of eigenfunctions\\ \hline
        $T$ & $3 \times 10^{-1}$~$\si{\micro\second}$ & Sampling interval\\ \hline
    \end{tabular}\vspace*{-6 mm}
    \label{tab:sim_params}
\end{table}

\vspace*{-0.2mm}
\subsection{Impact of Receptor Saturation on Steady-State}\label{sec:results:steady_state}

First, we consider the instantaneous release of \acp{NT} in the absence of enzymatic degradation, i.e., $\kappa_e C_E=0$.
In this case, there is no clearance mechanism in the synapse and, hence, the steady-state concentration of \acp{NT} in the synaptic cleft following a single release is non-zero.
The expected number of bound molecules without receptor saturation in the steady-state is \cite[Eq.~(17)]{lotter20a}
\begin{equation}
\frac{N \kappa_a}{\kappa_a + a \kappa_d}.\label{eq:steady_state}
\end{equation}

\begin{figure}[!t]
    \centering
    \includegraphics[width=.42\textwidth]{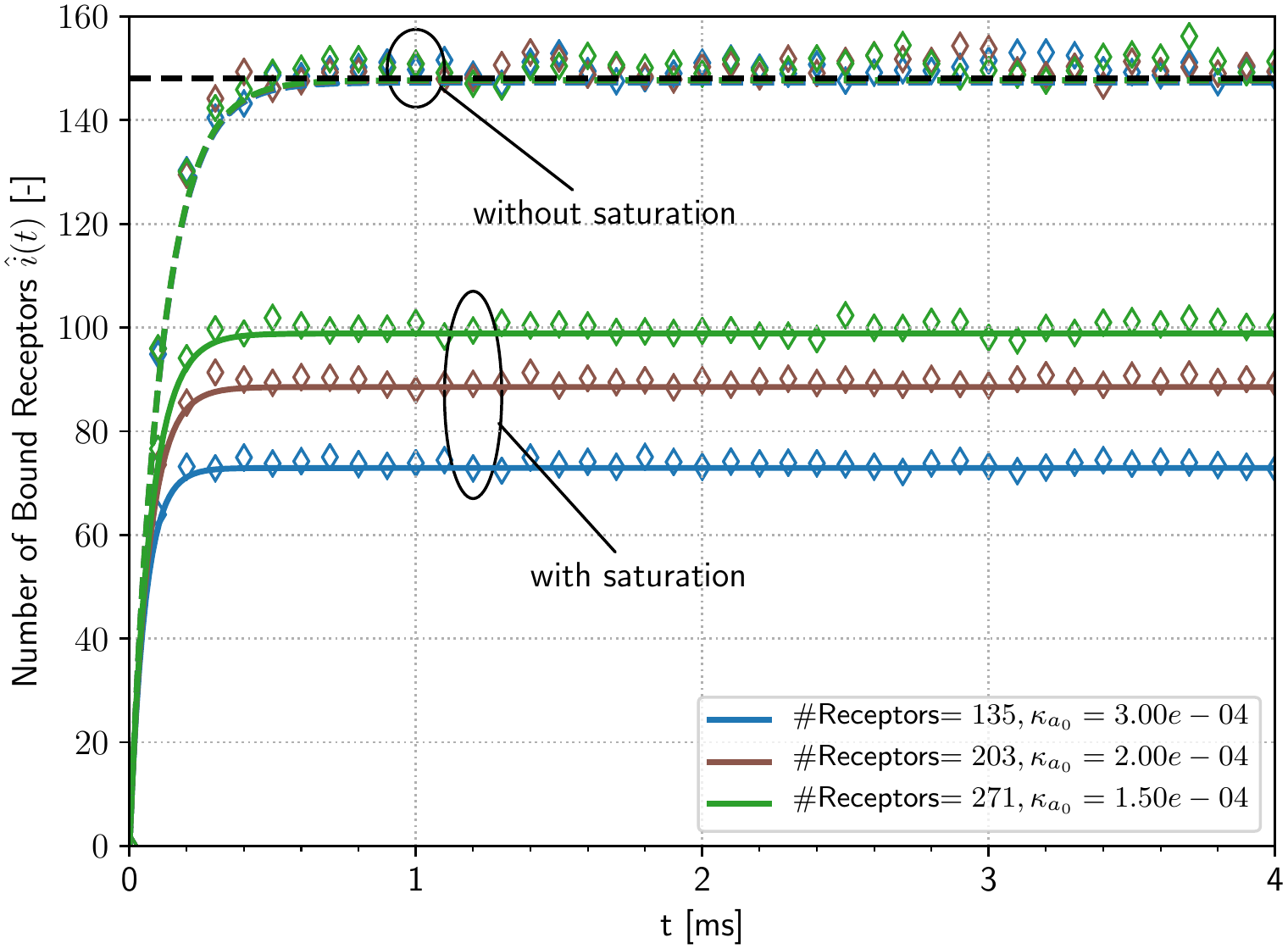}
    \vspace{-2mm}
    \caption{Results from particle-based simulation (diamond markers) and \eqref{eq:tf:13} (lines) for different numbers of postsynaptic receptors $C^*$ with (solid lines) and without (dashed lines) saturation when enzymes are absent. The disk adsorption probability $\kappa_{a_0}$ is adjusted such that the effective adsorption rate $\kappa_a$ remains constant. The black line shows the theoretical value from \eqref{eq:steady_state}.}\vspace*{-4 mm}
    \label{fig:no_deg_sat_no_sat}
\end{figure}

The results from the \ac{SSD} model from Section~\ref{sec:state-space_description}, i.e., \eqref{eq:tf:13}, and from the \ac{PBS} are shown in Fig.~\ref{fig:no_deg_sat_no_sat} for different numbers of postsynaptic receptors $C^*$ and different intrinsic binding rates $\kad$ (keeping the effective adsorption rate $\ka$ constant).
In the absence of saturation, we observe from Fig.~\ref{fig:no_deg_sat_no_sat} that both, \ac{SSD} and \ac{PBS}, agree with \eqref{eq:steady_state}, hence confirming the accuracy of the proposed model in the absence of saturation.
The small discrepancy between \ac{SSD} and \ac{PBS} is attributed to the use of boundary homogenization, cf.~A\ref{ass:bdr_hgz} in Section~\ref{sec:system_model:geometry_and_assumptions}.

Further, we observe from Fig.~\ref{fig:no_deg_sat_no_sat} that, in the presence of saturation, the steady-state number of bound molecules depends also on the number of available receptors.
Furthermore, we observe that the proposed \ac{SSD} and the \ac{PBS} agree very well also in the presence of saturation.

\vspace*{-0.2mm}
\subsection{Impact of Receptor Saturation on Single Release}\label{sec:results:single_release}

\begin{figure}[!t]
    \centering
    \includegraphics[width=.42\textwidth]{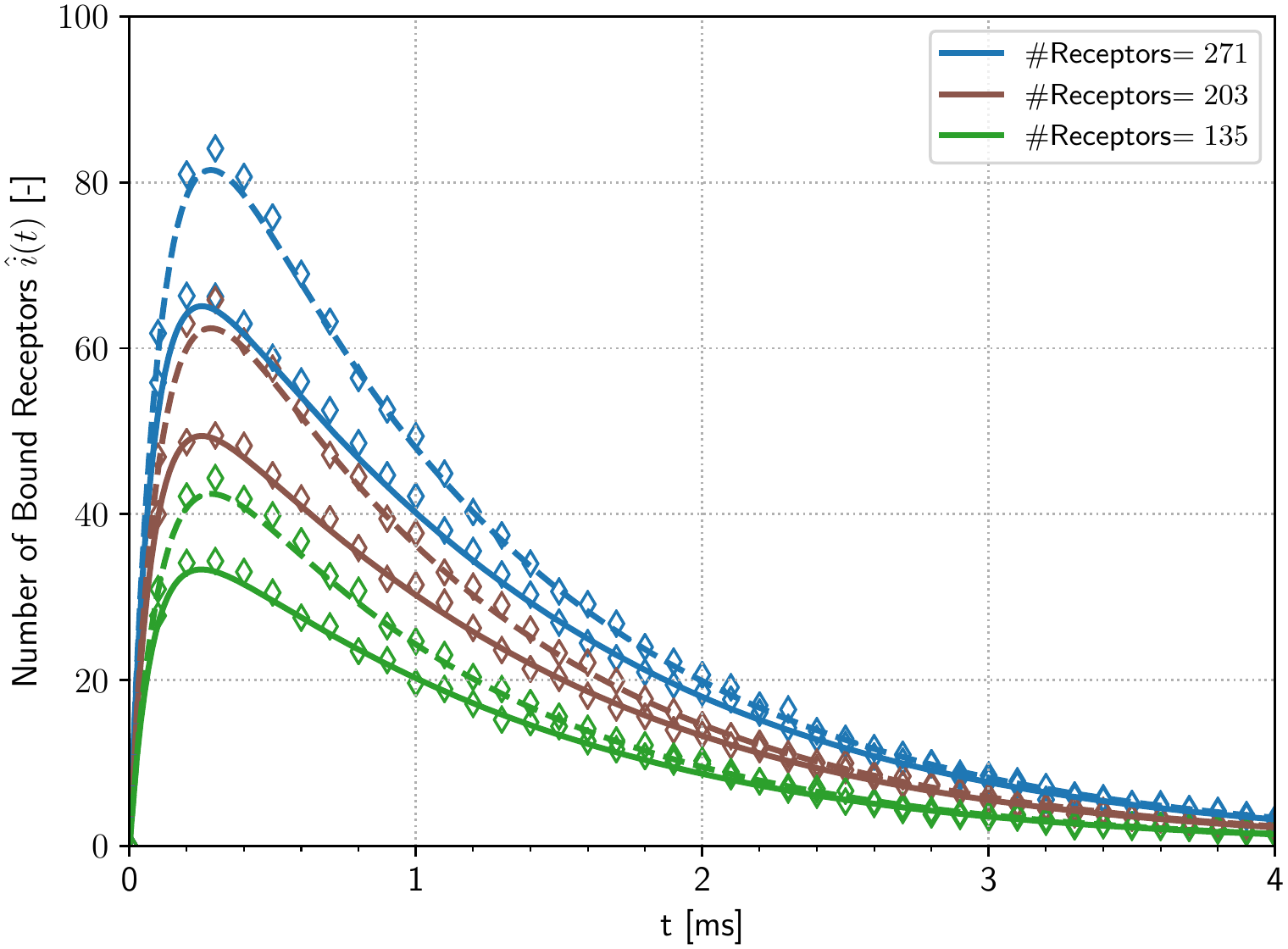}
    \vspace{-2mm}
    \caption{Results from particle-based simulation (diamond markers) and \eqref{eq:tf:13} (lines) for different numbers of postsynaptic receptors $C^*$ with (solid lines) and without (dashed lines) saturation when enzymes are present.}\vspace*{-6 mm}
    \label{fig:complete_sat_no_sat}
\end{figure}

Next, we investigate the impact of receptor saturation for one single instantaneous release of \acp{NT} when enzymes are present.
The results from the \ac{SSD} presented in Section~\ref{sec:state-space_description} and from \ac{PBS} are shown in Fig.~\ref{fig:complete_sat_no_sat} for different numbers of postsynaptic receptors $C^*$ and constant individual binding rates $\kad$.
In the presence of enzymatic degradation, all released molecules are eventually degraded.
We observe from Fig.~\ref{fig:complete_sat_no_sat} that receptor saturation significantly reduces the peak of the postsynaptic signal as compared to the configuration without receptor saturation.
Moreover, we also observe that the peak values in both cases, with and without receptor saturation, scale approximately linearly with the number of postsynaptic receptors in the considered regime.
Furthermore, we note that the effect of receptor saturation is most prominent around the peak occupancy at $t\approx 0.3$~$\si{\milli\second}$.
When fewer receptors are occupied, i.e., for $t>1.5$~$\si{\milli\second}$, the impact of receptor saturation on the postsynaptic signal is almost negligible.

\subsection{Impact of Receptor Saturation on Multiple Releases}\label{sec:results:multiple_releases}

\begin{figure}[!t]
    \centering
    \includegraphics[width=.42\textwidth]{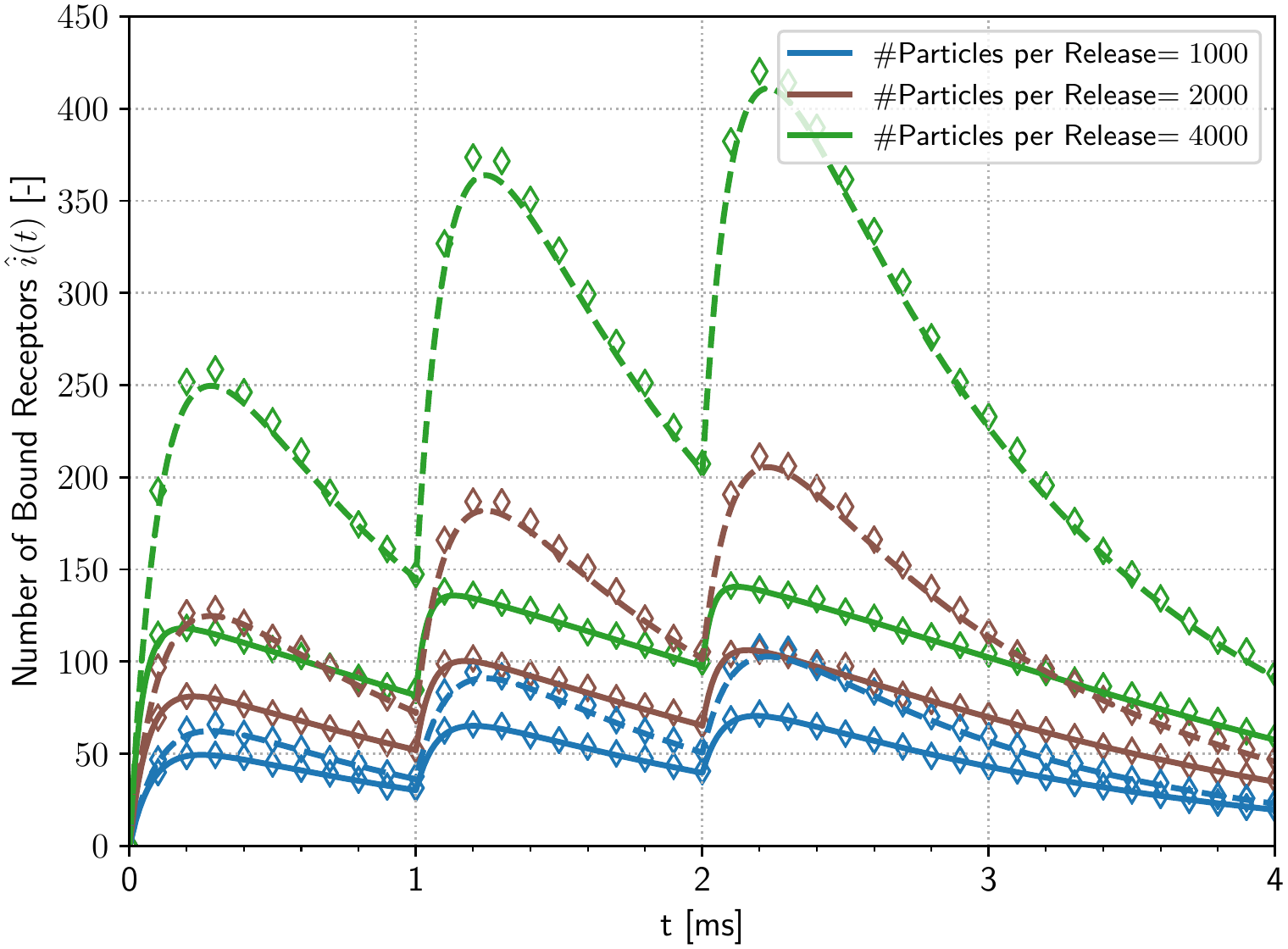}
    \vspace{-2mm}
    \caption{Results from particle-based simulation (diamond markers) and \eqref{eq:tf:13} (lines) for different number of released molecules $N$ with (solid lines) and without (dashed lines) saturation when enzymes are present.}\vspace*{-6 mm}
    \label{fig:multiple_release}
\end{figure}

Finally, we investigate the impact of receptor saturation on the postsynaptic signal for multiple \ac{NT} releases when enzymes are present.
The results for the \ac{SSD} presented in Section~\ref{sec:state-space_description} and \ac{PBS} are shown in Fig.~\ref{fig:multiple_release} for different numbers of released molecules $N$.
First, we observe that with receptor saturation, the peaks of the postsynaptic signals do not scale linearly with $N$.
Next, we observe that the impact of receptor saturation in terms of the peak values becomes more pronounced compared to the system without saturation as $N$ increases.
Finally, we note that, due to \ac{ISI}, for each $N$ the peak value following the second release of \acp{NT} (at $t \approx \SI{1.2}{\milli\second}$) is larger than the peak value following the first release.
Now, interestingly, this effect is significantly less pronounced in the presence of saturation.
In fact, this observation is consistent with experimental observations \cite{foster05} and has two reasons.
First, limiting the number of receptors naturally damps the signal because fewer receptors are available.
Second, as the number of receptors decreases, fewer molecules are bound simultaneously and, consequently, molecules become more exposed to degradation and the channel is cleared faster.

\vspace{-0.5mm}
\section{Conclusions}\label{sec:conclusion}
In this work, we have presented a novel model for synaptic \ac{MC} in the presence of enzymatic degradation and receptor saturation.
We have shown that the proposed deterministic model is consistent with stochastic \ac{PBS} and produces biologically plausible results.
The proposed state-space description allows for the numerically stable and efficient computation of the solution of the proposed model.

While the presented model incorporates enzymatic degradation, it would be interesting to also consider other channel clearance mechanisms such as presynaptic uptake or molecule uptake at glial cells.
However, due to space constraints, this is left for future work.

\vspace*{-2.35mm}
\bibliographystyle{IEEEtran}
\bibliography{IEEEabrv,%
        icc21_sebastian_bib,%
		icc21_max_bib%
		}

\end{document}